\documentclass[12pt,a4paper,superscriptaddress,pre,showpacs]{revtex4}

\usepackage{graphicx}
\usepackage{amssymb}

\newcommand{\EQ}{Eq.~}
\newcommand{\EQS}{Eqs.~}
\newcommand{\FIG}{Fig.~}
\newcommand{\FIGS}{Figs.~}
\newcommand{\TAB}{Tab.~}
\newcommand{\SEC}{Sec.~}
\newcommand{\SECS}{Secs.~}

\begin{document}

%\begin{frontmatter}

\title{Multi-state Epidemic Processes on Complex Networks}
\author{Naoki Masuda}
\address{Laboratory for
Mathematical Neuroscience, RIKEN Brain Science Institute, 2-1,
Hirosawa, Wako, Saitama 351-0198, Japan}
%\ead{masuda@brain.riken.jp}
\author{Norio Konno}
\address{Faculty of Engineering, Yokohama National University,
79-5, Tokiwadai, Hodogaya, Yokohama 240-8501, Japan}
%\ead{norio@mathlab.sci.ynu.ac.jp} \corauth[cor]{Corresponding
%author. Address: Laboratory for Mathematical Neuroscience, RIKEN Brain
%Science Institute, 2-1, Hirosawa, Wako, Saitama 351-0198, Japan,
%Phone:+81-48-467-9664, FAX:+81-48-467-9693}
%\newpage

\begin{abstract}
{\bf Abstract}:
Infectious diseases are practically represented by 
models with multiple states and
complex transition rules corresponding to, for example, birth, death,
infection, recovery, disease progression, and quarantine. In addition,
networks underlying infection events are often much more complex than
described by meanfield equations or regular lattices. In models with
simple transition rules such as the SIS and SIR models, heterogeneous
contact rates are known to decrease epidemic thresholds.  We analyze
steady states of various multi-state disease propagation models with
heterogeneous contact rates.  In many models, heterogeneity simply
decreases epidemic thresholds. However, in models with
competing pathogens and mutation, coexistence of different pathogens
for small infection rates requires network-independent conditions in
addition to heterogeneity in contact rates.
Furthermore, models without spontaneous neighbor-independent
state transitions, such as
cyclically competing species, do not show heterogeneity effects.
\end{abstract}

%\begin{keyword}
%epidemiology \sep complex networks \sep
%scale-free networks \sep contact process \sep epidemic threshold

%\PACS 89.75.Hc \sep 87.19.Xx \sep 87.23.Ge \sep 89.75.Da
%\end{keyword}
%\end{frontmatter}

\maketitle
%\newpage

\section{Introduction}\label{sec:introduction}

Many epidemic systems can be represented as a graph where vertices
stand for individuals and an edge connecting a pair of vertices
indicates interaction between individuals.
Since late 1990s, real networks underlying disease transmission
have been recognized to be
complex and not approximated by conventional graphs such as lattices,
regular trees, or classical random graphs.  Not only being complex,
many networks have the small-world and scale-free properties
\cite{Albert02,NewmanSIAM}.  An important property of 
small-world networks is
that the distance between a pair of vertices, or the minimum
number of edges connecting two vertices, is fairly small on average.
The scale-free property means that
 the vertex degree $k$, or the number of contacts with 
other individuals (vertices) for a given individual,
has a power law distribution:
$p_k\propto k^{-\gamma}$ \cite{Barabasi99}.  Typically,
$\gamma$ ranges between 2 and 3 \cite{Albert02,NewmanSIAM}. In
contrast, the degree is uniform on, for example,
the square lattice ($k=4$).

Extending earlier data on heterogeneous sexual contact rates of humans
\cite{Hethcote84,Anderson86,May88TRANS}, current real data
support that scale-free networks underly propagation of sexually
transmitted diseases such as gonorrhea, genital herpes, HIV
\cite{Colgate,Andersonbook,Liljeros,Schneeberger}, and computer
viruses \cite{Pastor01PRL,Pastorbook,NewmanSIAM}. Theoretically,
epidemic thresholds above which epidemic outbreaks or endemic states
ensue scale as $\left<k\right> / \left<k^2\right>$. Here
$\left< k\right> \equiv \sum_{k=1}^{\infty} k p_k$ is the mean degree,
and $\left< k^2\right> \equiv \sum_{k=1}^{\infty} k^2 p_k$ is the
second moment of the degree.
Then a
heterogeneous population, which has a large $\left<k^2\right>$ in
comparison with $\left<k\right>$, yields a smaller epidemic threshold
and are more likely to elicit outbreaks.
Particularly, scale-free
networks with $\gamma\le 3$ have epidemic thresholds equal to 0
because $\left<k^2\right>$ diverges. Then, disease
propagation on a global scale can occur for an arbitrary small
infection rate.  This holds for the percolation
\cite{Albert00NAT,Callaway00,Cohen00PRL,Newman02_bond}, the contact process (also
called the SIS model) that describes endemic diseases such as
gonorrhea \cite{Hethcote84,Pastor01PRL,Pastor01PRE},
and the SIR model that describes single-shot
outbreak dynamics such as HIV
\cite{Anderson86,May88TRANS,May01,Moreno02EPB}.

\begin{figure}
\begin{center}
\includegraphics[height=2.25in,width=2.25in]{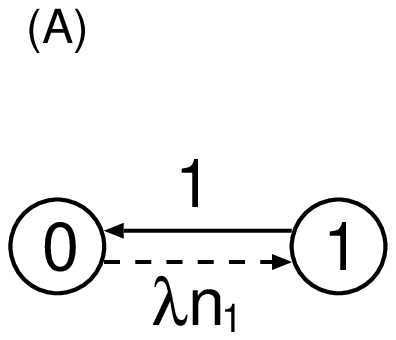}
\includegraphics[height=2.25in,width=2.25in]{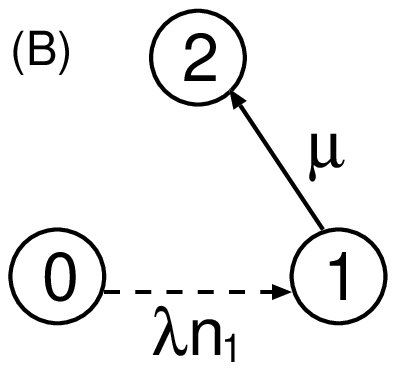}
\includegraphics[height=2.25in,width=2.25in]{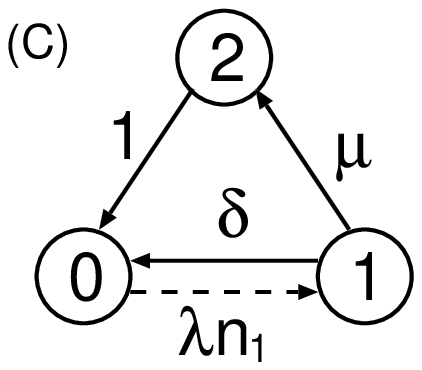}
\caption{The state-transition rules of (A) 
the contact process, (B) the SIR model, and (C) the SIRS model.
Solid lines represent transitions independent
of states of neighbors. Dashed lines represent
neighbor-dependent transitions. The values indicate the
transmission rates, where $n_i$ is the number of
neighbors with state $i$.}
\label{fig:known}
\end{center}
\end{figure}

Apart from heterogeneous contact rates, realistic epidemic
processes are also more complex than merely described by the percolation,
the contact process, or the SIR model. Practical epidemic models often
have more detailed states corresponding to, for example, exposed,
quarantine, and mutated.  Accordingly, state-transition rules
become more complicated
\cite{Hethcote84,May88TRANS,Andersonbook,Hopbauerbook}.
It is convenient to represent the transition rules of an epidemic model
by a schematic figure like \FIG\ref{fig:known}.  Solid lines
mean state-transition routes whose event rates are
independent of the states of neighbors (e.g. death, recovery,
mutation).  Dashed lines represent ones whose rates are proportional
to the neighbors' states (e.g. reproduction, infection).
Then, the contact process and the SIR model are represented by
\FIG\ref{fig:known}(A) and \ref{fig:known}(B), respectively.
For example, in the contact process, 
a susceptible individual (state 0) turns infected (state 1) at a rate
proportional to the number of infected individuals in the
neighborhood $(=n_1)$. The infection rate is denoted by $\lambda$, and
the spontaneous recovery rate is set equal to unity.

Understanding of multi-state contagion
models on complex networks may be helpful
in combatting real epidemics.  In this direction,
Liu and coworkers analyzed the SIRS model,
which has the state of partial immunity R and is schematically shown
in \FIG\ref{fig:known}(C) \cite{Liu04JSM}. They also studied
the household model, in which each vertex has
a graded state corresponding to the number of patients in a household
\cite{Liu04PHA}.
They showed that the epidemic thresholds
are proportional to $\left<k\right>/\left<k^2\right>$ also
in these models. However, how general the universality of the
$\left<k\right>/\left<k^2\right>$-scaling is has not been addressed.

In this paper, we analyze various
multi-state epidemic processes originally proposed for lattices or
populations with homogeneous contact rates. We concentrate on
endemic models in which the phase boundary between the coexistence
state, where both susceptible and infected individuals can survive, and the
disease-free state with the susceptible individuals 
only is concerned. Of course,
we then disregard the other important class of epidemic models
including the SIR model, in which
possibility of one-shot waves of outbreaks rather than
coexistence is concerned
(for this distinction, see \cite{Hethcote84,Andersonbook}).
Instead of covering both classes,
we investigate phase boundaries of endemic models with more than two
states and possibly with more than two phases.
We show that the critical
infection rates are largely extinct as $\left<k^2\right>\to\infty$.
However, there are
a couple of important exceptions to this rule: models with mutation or
cyclic interaction. In
\SEC\ref{sec:single}, we start with studying two models that follow the
$\left<k\right>/\left<k^2\right>$-scaling.
Sections~\ref{sec:two_pathogens}, \ref{sec:double}, \ref{sec:rock},
and \ref{sec:malaria} are devoted to the analysis of representative
models with different characteristics and outcomes.  Some of them
do not follow the $\left<k\right>/\left<k^2\right>$-scaling.
The results are summarized and discussed in
\SEC\ref{sec:conclusions}.

\section{Models with Single Neighbor-dependent Transitions}\label{sec:single}
\begin{figure}
\begin{center}
\includegraphics[height=2.25in,width=2.25in]{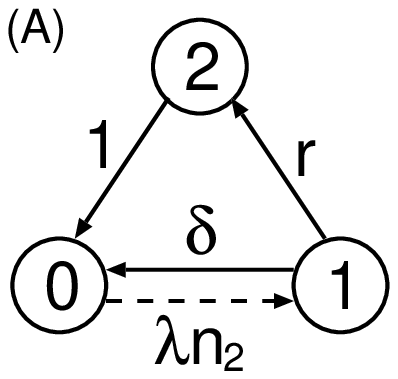}
\includegraphics[height=2.25in,width=2.25in]{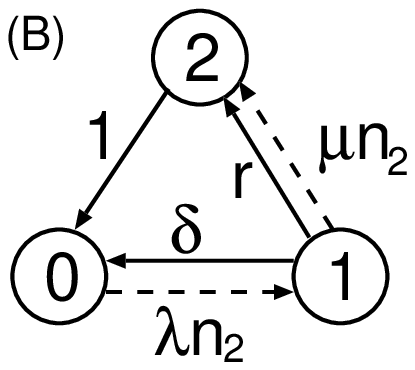}
\caption{The state-transition rules of (A) the two-stage contact process
and (B) the tuberculosis model.}
\label{fig:twostage}
\end{center}
\end{figure}

\begin{figure}
\begin{center}
\includegraphics[height=2.25in,width=2.25in]{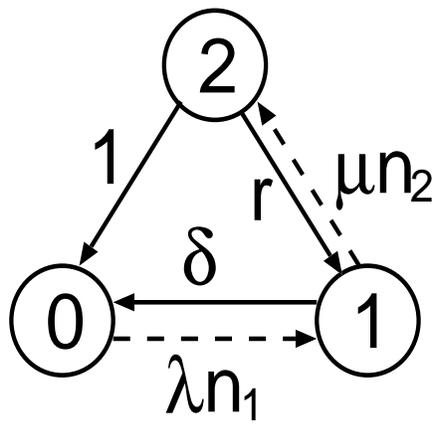}
\caption{The state-transition rules of the double infection
model.}
\label{fig:double}
\end{center}
\end{figure}

In the following, we investigate various contagion models with
different transition rules. Because we can never be exhaustive, our
strategy is to study known multiple-state models.
It will turn out that the extinction of epidemic
thresholds as $\left<k^2\right>/\left<k\right>$ goes to infinity is
not a universal story, as summarized in
\SEC\ref{sec:conclusions}. What determines
the nature of phase boundaries is considered to be the
gross organization of the transition pathways, some of which are
independent of the state of the neighbors (solid lines in
\FIGS\ref{fig:known}, \ref{fig:twostage}, \ref{fig:twoout},
\ref{fig:double}, \ref{fig:rock_voter}) and others are 
neighbor-dependent (dashed lines).
For the introductory purpose, we analyze in this section two
three-state models whose epidemic thresholds are proportional to
$\left<k\right>/\left<k^2\right>$.

\subsection{Two-stage Contact Process}\label{sub:twostage}

{\bf Model}

The state-transition
rules of the two-stage contact process \cite{Krone} are
schematically shown in \FIG\ref{fig:twostage}(A).  It differs from the
SIRS model (\FIG\ref{fig:known}(C)) in that the transition rate from 0 to
1 is proportional to the density of the neighbors with state 2 (not 1)
of a state-0 vertex. This model corresponds to dynamics with two life
stages.  The states can be
interpreted as 0: vacant, 1: occupied by young individuals, and 2:
occupied by adults.  Only adults are reproductive
and generate offsprings
in neighboring vacant sites at a birth rate equal to $\lambda$. In
other words, a birth event occurs at 
a vacant site at a rate proportional to
$\lambda$ and the number of neighboring adults.
Youngs (state 1) spend random time of mean
$r$ before becoming adult (state 2).  They are also subject to random
death events at a rate of $\delta$.  Adults die at a rate of 1, which
gives normalization of the entire model.
Alternatively, we can interpret the three states
as 0: vacant, 1: partially occupied, and 2: fully occupied colonies. Then,
only fully occupied colonies are potent enough to colonize vacant
lands. In the disease analogue, 0 is susceptible, 1 is infected but in
the incubation period (not infectious), and 2 is infected
and infectious. We note that the model corresponds to the SEIRS model
without the recovery state R.

The ordinary contact process (\FIG\ref{fig:known}(A)) is recovered as
$r=\infty$. In this case, a state-1 vertex
immediately turns into state 2.  The meanfield approximation and the
rigorous results on regular lattices conclude that there are two
phases:
$\{0\}$ (the steady
state that consists only of state 0) 
and $\{0,
1, 2\}$ (positive probability of steady coexistence of 0, 1, and 2).
Naturally, large $\lambda$ or $r$ promotes survival
\cite{Krone}.  Particularly, in the meanfield approximation, these two
phases are divided by $r\lambda = \delta + r$.

{\bf Analysis}

To derive meanfield dynamics for populations with heterogeneous
contact rates,
let us denote by $p_k$ the probability that a vertex has degree $k$.
Obviously, $\sum_{k=1}^{\infty} p_k = 1$.
We denote by $\rho_{i,k}$ ($i=1, 2$)
the probability that a vertex with degree $k$ takes
state $i$. The probability that a vertex with degree $k$ takes
state 0 is equal to $1-\rho_{1,k}-\rho_{2,k}$.
The probability that a neighbor located at the end of a 
randomly chosen edge takes state $i$ is denoted by
$\Theta_i$. This edge-conditioned 
probability, particularly $\Theta_2$ in this model,
is used to determine the effective 
infection rate \cite{Pastor01PRL,Pastor01PRE}.

The dynamics are given by
\begin{eqnarray}
\dot{\rho}_{1,k} &=& \lambda (1-\rho_{1,k}-\rho_{2,k}) k
 \Theta_2 - (\delta+r) \rho_{1,k},\nonumber\\
\dot{\rho}_{2,k} &=& r \rho_{1,k} - \rho_{2,k}.
\label{eq:twostage1}
\end{eqnarray}
We note that the effective
birth rate (the first term in the first equation of
\EQ(\ref{eq:twostage1})) is proportional to
$k\Theta_2$, which is the average number of state-2 vertices in the
neighborhood of a degree-$k$ vertex.

When we choose an arbitrary edge,
the probability that 
a specific vertex is connected to this edge
is proportional to its degree $k$ 
\cite{Callaway00,Cohen00PRL,Pastor01PRL,Pastor01PRE,Newman02_bond}.
Therefore, we obtain
\begin{equation}
\Theta_i = \frac{\sum_k k p_k \rho_{i,k}(t)}{\left< k\right>}.
\label{eq:pastor2}
\end{equation}

The steady state is given by
\begin{eqnarray}
\rho^*_{1,k} &=& \frac{\rho^*_{2,k}}{r},\nonumber\\
\rho^*_{2,k} &=& \frac{\lambda r \Theta^*_2 k}
{\delta+r + \lambda(r+1) \Theta^*_2 k},
\label{eq:twostage2}
\end{eqnarray}
where $*$ indicates the steady state.
Plugging \EQ(\ref{eq:twostage2}) into \EQ(\ref{eq:pastor2}) leads to
\begin{equation}
\Theta^*_2 = \frac{1}{\left< k\right>}\sum_k
\frac{\lambda r \Theta^*_2 k^2 p_k}
{\delta+r+\lambda(r+1)\Theta^*_2 k}.
\label{eq:twostage3}
\end{equation}
Equation~(\ref{eq:twostage3}) is satisfied when
$\Theta^*_2 = 0$, corresponding
to the disease-free state $\{0\}$.
When $0<\Theta^*_2 < 1$, state 2 survives.
Equation~(\ref{eq:twostage2}) implies
$\rho^*_{1,k}, \rho^*_{2,k}>0$ in this situation.
Accordingly, $\Theta^*_2>0$ is
equivalent to the $\{0, 1, 2\}$ phase.
Because the RHS of \EQ(\ref{eq:twostage3}) is
smaller than the LHS when $\Theta^*_2 = 1$, 
the condition for the 
$\{0, 1, 2\}$ phase is
\begin{equation}
\left. \frac{\partial}{\partial\Theta^*_2}
\left( \frac{1}{\left< k\right>}\sum_k
\frac{\lambda r \Theta^*_2 k^2 p_k}
{\delta+r+\lambda(r+1)\Theta^*_2 k}.
\right)
\right|_{\Theta^*_2=0}
> 1.
\label{eq:twostage_derivative_cnd}
\end{equation}
This yields
\begin{equation}
\lambda > \frac{(\delta+r)\left<k\right>}{r \left<k^2\right>}.
\label{eq:twostage4}
\end{equation}
When $p_k\propto k^{-\gamma}$ ($\gamma\le 3$),
the critical infection rate
vanishes. This agrees with the results for the percolation,
the contact process, and the SIR model.

\subsection{Tuberculosis Model}\label{sub:tuber}

{\bf Model}

A next example is the tuberculosis model shown in
\FIG\ref{fig:twostage}(B) \cite{Schinazi03_JTB}.  In the tuberculosis,
a majority of the infected individuals does not become infectious
before recovery. Accordingly, state 0 corresponds to healthy, 1
to infected but not infectious, and 2 to infected and infectious.
This model, which is an extension of the two-stage contact
process, has two important features.  One is that we have two
state-transition routes that depend on neighbors' states.  Both of them
depend on the density of infectious individuals (state 2). Antoher
feature is that disease progression from state 1 to 2 is possible
through two parallel routes. An early-stage patient (state 1) can
develop the disease on its own at a rate of $r$. At the same time,
infectious patients (state 2) in the neighborhood
increases the possibility of disease
progression for a state-1 individual.

When $r=0$, the $\{0, 1, 2\}$ phase is possible on regular lattices
but impossible in perfectly mixed populations
\cite{Schinazi03_JTB}.
When $r>0$, the $\{0\}$ phase and the $\{0, 1, 2 \}$ phase
appear for both lattices and meanfield populations, depending on
values of other parameters. According to the ordinary meanfield analysis,
these phases are divided by $\lambda r = (\delta + r)$,
which is the condition identical to that for the two-stage contact
process. 

{\bf Analysis}

In heterogeneous populations, the dynamics become
\begin{eqnarray}
\dot{\rho}_{1,k} &=& \lambda(1-\rho_{1,k}-\rho_{2,k}) k
\Theta_2 -\mu \rho_{1,k} k\Theta_2
- (\delta+r)\rho_{1,k},\nonumber\\
\dot{\rho}_{2,k} &=& \mu \rho_{1,k} k
\Theta_2 +r \rho_{1,k} - \rho_{2,k}.
\label{eq:schinazi03_jtb_a}
\end{eqnarray}
The steady state is given by
\begin{equation}
\left(\begin{array}{c}
\rho^*_{1,k}\\ \rho^*_{2,k}
\end{array}\right)
= \frac{\lambda\Theta^*_2 k}
{\delta+r + (\lambda+\mu+\lambda r)\Theta^*_2 k
+ \lambda\mu\Theta^{* 2}_2 k^2}
\left(\begin{array}{c}
1\\ r + \mu k\Theta^*_2
\end{array}\right),
\label{eq:schinazi03_jtb_b}
\end{equation}
and
\begin{equation}
\Theta^*_2 = \frac{1}{\left< k\right>}
\sum_k 
\frac{\lambda k^2 (r+\mu k\Theta^*_2)\Theta^*_2 p_k}
{\delta+r + (\lambda+\mu+\lambda r) k\Theta^*_2
+ \lambda\mu k^2\Theta^{* 2}_2}.
\label{eq:schinazi03_jtb_c}
\end{equation}
The $\{0, 1, 2\}$ phase appears when
$\Theta^*_2>0$, or
\begin{equation}
\lambda > \frac{(\delta+r)\left<k\right>}{r \left<k^2\right>}.
\end{equation}
This is identical to \EQ(\ref{eq:twostage4}).

\section{Models with Competing Pathogens}\label{sec:two_pathogens}

In many epidemics, multiple strains or pathogens with different
transmissivity, virulence, and mobility coexist. They compete with
each other by preying on common susceptible individuals.  On top of
that, transitions between infected states with different pathogens or
stages of the disease can occur owing to mutation and disease
progression.  Coexistence of two distinct diseases in a population also
introduces competition between them.

In terms of schematic diagrams, this means that multiple
neighbor-dependent infection pathways (dashed lines) emanate from a
susceptible state.
We analyze three models with competiting pathogens and show
that epidemic thresholds are not entirely governed by the
degree-dependent factors in two models.

\subsection{A Minimal Model with Pathogen Competition and Mutation}\label{sub:example1}

{\bf Model}

Let us consider a minimal model for competing pathogens 
whose transition rules are
shown in \FIG\ref{fig:twoout}(A). State 1 and state 2
correspond to patients infected with strain 1 and strain 2, respectively. 
The two strains compete with each other by 
feeding on the common resource, namely, 
susceptible individuals (state 0), at rates $\beta_1$ and $\beta_2$.
Any patient spontaneously recovers at rate 1.
Unidirectional and spontaneous mutations occur (1 $\to$ 2) at rate
$r$ \cite{Schinazi01_MPRF}. As the mutation rate increases, 
strain 2 becomes more viable than strain 1.
In the poupulation ecology context, 
the three states can be also interpreted as
empty (0), species A (1), and species B (2).

Exploiting the fact that the absence of state-1 vertices is equivalent
to the
contact process, Schinazi showed for regular lattices that there are
three phases $\{0\}$, $\{0, 2\}$, and $\{0, 1, 2\}$ separated by
nontrivial critical lines \cite{Schinazi01_MPRF}.  Small $r$ and large
$\beta_1$ support the $\{0, 1, 2\}$ state. The meanfield
analysis predicts that 
the boundary between $\{0\}$ and $\{0, 2\}$ is $\beta_2 =
1$, and $\{0, 1, 2\}$ appears when both $\beta_1
> \beta_2 (r+1)$ and $\beta_1>r+1$ are additionally satisfied.

{\bf Analysis}

For heterogeneous populations, the dynamics are given by
\begin{eqnarray}
\dot{\rho}_{1,k} &=& \beta_1 \left( 1-\rho_{1,k}-\rho_{2,k} \right)
k \Theta_1 - (r+1) \rho_{1,k},\nonumber\\
\dot{\rho}_{2,k} &=& \beta_2 \left( 1-\rho_{1,k}-\rho_{2,k}\right)
k \Theta_2 + r\rho_{1,k} - \rho_{2,k}.
\label{eq:twoout1}
\end{eqnarray}
The boundary between $\{0\}$ and $\{0,2\}$ is obtained easily.
Application of the contact-process result
implies that $\{0, 2\}$ emerges 
when
\begin{equation}
\beta_2 > \frac{\left<k\right>}{\left<k^2\right>},
\end{equation}
which extends the condition $\beta_2>1$ for homogeneous populations.

We next analyze the boundary between $\{0,2\}$ and $\{0,1,2\}$.
With the steady state
\begin{equation}
\left(\begin{array}{c}
\rho^*_{1,k}\\ \rho^*_{2,k}
\end{array} \right) =
\frac{k}{r+1+\left(r+1\right)
\left(\beta_2\Theta^*_2+\beta_1\Theta^*_1\right)k}
\left(\begin{array}{c}
\beta_1 \Theta^*_1\\ \beta_1 r\Theta^*_1
+\beta_2(r+1)\Theta^*_2
\end{array} \right),
\label{eq:twoout2}
\end{equation}
we obtain
\begin{equation}
\Theta^*_1 = \frac{\beta_1 - \beta_2(r+1)}
{r\beta_1}\Theta^*_2 \equiv C_1 \Theta^*_2,
\label{eq:twoout3}
\end{equation}
and
\begin{equation}
\Theta^*_2 = \frac{1}{\left< k\right>\left(r+1\right)}
\sum_k \frac{\beta_1\Theta^*_2 k^2 p_k}
{1+\left(\beta_1 C_1 + \beta_2\right)\Theta^*_2 k}.
\end{equation}
In the $\{0, 1, 2\}$ phase, $\Theta_1>0$ and $\Theta_2>0$ are
simultaneously satisfied.
The condition $\Theta^*_2>0$ is satisfied when
\begin{equation}
\beta_1 > \frac{(r+1)\left<k\right>}{\left<k^2\right>}.
\label{eq:twoout-012}
\end{equation}
Equation~(\ref{eq:twoout-012})
extends one of the conditions for the $\{0,1,2\}$ phase 
in the homogeneous population: $\beta_1>r+1$. 
Imposing $\Theta^*_1>0$ leads to
$\beta_1 > \beta_2(r+1)$, a condition 
independent of heterogeneity in contact rates.
Because of this condition,
$\left<k^2\right>\to\infty$ means survival
of state 2 but not necessarily survival of state 1.
Increasing 
$\left<k^2\right>/\left<k\right>$ makes 
the survival of state 1 and state 2 more or less equally likely
because they feed on the susceptibles in the same manner.
Then, the strengths of 1 and 2 must be balanced so that 
one does not completely devour the other. 
This constrains the range of the mutation rate in a degree-independent
manner.

Figure~\ref{fig:pathogen} shows numerically obtained steady densities
of state 1 (A) and 2 (B) in networks with 10000 vertices and $\left<
k\right>=12$.  Initially, each vertex takes one of the three states
independently with probability $1/3$.  We set $\beta_1 = 0.5$ and
$\beta_2 = 0.25$, which predicts the phase boundary:
$r=\beta_1/\beta_2 - 1 = 1$.

Scale-free networks with the degree distribution
$p_k\propto k^{\gamma}$ with $\gamma=2.5$
(thickest solid lines), $\gamma=3.0$ (moderate solid lines),
and $\gamma=4.0$ (thinnest solid
lines) are produced using a static model
\cite{Goh01PRL}. Even though produced networks can be disconnected in
general, more than 95 \% of the vertices constitute one component in
every run of our simulations.
The results for the random graph, whose $p_k$ follows the Poisson
distribution, are also shown (dashed lines).
Regardless of networks, the numerical results support
the existence of finite thresholds in terms of
$r$. The values of critical $r$ 
are slightly smaller than $r=1$ but
do not vary so much
for different $p_k$, as our theory predicts. 

\subsection{Model of Drug-resistant Diseases}\label{sub:example2}

{\bf Model} 

In many diseases such as tuberculosis, drug-resistant strains emerge
by the mutation of 
a wild strain \cite{Blower}.
Misuse of antibiotics is a major cause of appearance of such stronger
strains. 
A four-state model shown in \FIG\ref{fig:twoout}(B)
represents spreads of a drug-resistant strain
\cite{Schinazi99_JSP}. The states 0, 1, 2,
and 3 represent empty, susceptible, infected with the wild strain, and
infected with the drug-resistant strain, respectively. 
The wild strain (state 2) and the drug-resistant strain (state 3)
compete with each other to devour the susceptible (state 1). Only the
drug-resistant strain is supposed to be fatal with death rate 1.
When infected by the wild strain, which is not lethal, an individual 
recovers at rate $r$ owing to the drug.
The mutation rate is denoted by $\phi$.

The model without state 2 (or $\beta_2 = 0$)
is equivalent to the SIRS model
in \FIG\ref{fig:known}(C) with $\delta=0$.
On regular lattices,
$\{0, 1, 2, 3\}$
appears when $r+\phi$ is small enough. Otherwise,
state 2 is extinguished \cite{Schinazi99_JSP}.
Based on the meanfield analysis,
the $\{0, 1, 2, 3\}$ phase ensues when
\begin{equation}
\phi>0,\quad \beta_2>r+\phi,\quad \frac{\beta_2}{r+\phi}>\beta_3.
\label{eq:schinazi99_jsp_mf1}
\end{equation}
If any of the conditions in 
\EQ(\ref{eq:schinazi99_jsp_mf1}) is violated and
$\beta_3>1$, then we have $\{0, 1, 3\}$. Otherwise,  
only state 1 survives (the $\{1\}$ phase).

{\bf Analysis}

In heterogeneous populations, we obtain
\begin{eqnarray}
\dot{\rho}_{1,k} &=& \beta_1 \left(
1-\rho_{1,k}-\rho_{2,k}-\rho_{3,k}\right) + r \rho_{2,k}
- \beta_2 \rho_{1,k} k \Theta_2
- \beta_3 \rho_{1,k} k \Theta_3,\nonumber\\
\dot{\rho}_{2,k} &=& \beta_2 \rho_{1,k} k \Theta_2
- (r+\phi) \rho_{2,k},\nonumber\\
\dot{\rho}_{3,k} &=& \beta_3 \rho_{1,k} k \Theta_3
+ \phi \rho_{2,k} - \rho_{3,k}.
\label{eq:schinai_jsp99a}
\end{eqnarray}
The steady densities are given by
\begin{equation}
\left(\begin{array}{c}
\rho^*_{1,k}\\ \rho^*_{2,k}\\ \rho^*_{3,k}
\end{array} \right) =
\frac{\beta_1}{\Delta}
\left(\begin{array}{c}
r+\phi\\
\beta_2 \Theta^*_2 k\\
\left[\beta_2\phi\Theta^*_2+
\beta_3\left(r+\phi\right) \Theta^*_3\right] k
\end{array} \right),
\label{eq:schinazi99_jsp_b}
\end{equation}
where
\begin{equation}
\Delta \equiv \beta_1(r+\phi)+\left[
\left(\beta_1 +\phi + \beta_1\phi \right)\beta_2\Theta^*_2
+ \left(\beta_1+1\right)\beta_3\left(r+\phi\right)\Theta^*_3
\right]k.
\label{eq:schinazi99_jsp_c}
\end{equation}
Equation~(\ref{eq:schinazi99_jsp_b}) assures that
$\rho^*_{1,k}>0$.
In addition, 
\begin{equation}
\Theta^*_3 = \frac{\beta_2 \phi}{\beta_2 - \beta_3(r+\phi)}
 \Theta^*_2
\label{eq:schinazi99_jsp_d}
\end{equation}
provided that $\beta_2 \big/ (r+\phi) > \beta_3$. Under this condition, 
$\Theta^*_2>0$ is equivalent to
the $\{0, 1, 2, 3\}$ phase. Combining
\EQS(\ref{eq:pastor2}),
(\ref{eq:schinazi99_jsp_b}) and 
(\ref{eq:schinazi99_jsp_d}) results in
\begin{equation}
\Theta^*_2 = \frac{1}{\left< k\right>}\sum_k
\frac{\beta_2 \Theta^*_2 k^2 p_k}{\Delta}.
\label{eq:schinazi99_jsp_e}
\end{equation}
Since $\Delta$ is of the form: $\Delta = C_3 + C_4 k\Theta^*_2$,  
\EQ(\ref{eq:schinazi99_jsp_e})
is essentially the same as \EQ(\ref{eq:twostage3}). Accordingly,
$\Theta^*_2>0$ when
\begin{equation}
\beta_2 > \frac{\beta_1(r+\phi)\left<k\right>}{\left<k^2\right>}.
\label{eq:schinazi99_jsp_f}
\end{equation}
Even for large $r+\phi$, sufficiently heterogeneous 
$p_k$ with large $\left<k^2\right>\big/\left<k\right>$ makes
\EQ(\ref{eq:schinazi99_jsp_f}) valid. However, for
existence of state 2, $\beta_2(r+\phi)>\beta_3$ must be also
satisfied. This condition is independent of the degree distribution
and identical to one of the conditions for the $\{0, 1, 2, 3\}$ phase
in the ordinary meanfield case
(see \EQ(\ref{eq:schinazi99_jsp_mf1})). As is the case for the
previous model (\SEC\ref{sub:example1}), this condition regulates the
rates of mutation and spontaneous recovery so that the strengths of
the wild strain and the drug-resistant strain are roughly balanced.

If $\beta_2 \big/ (r+\phi) \le \beta_3$, then $\Theta^*_2=0$. In this
case, \EQ(\ref{eq:schinazi99_jsp_b}) reduces to
\begin{equation}
\left(\begin{array}{c}
\rho^*_{1,k}\\ \rho^*_{3,k}
\end{array} \right) =
\frac{\beta_1}{\beta_1+ (\beta_1+1)\beta_3\Theta^*_3 k}
\left(\begin{array}{c}
1\\
\beta_3 \Theta^*_3 k
\end{array} \right),
\label{eq:schinazi99_jsp_g}
\end{equation}
which yields
\begin{equation}
\Theta^*_3 = \frac{1}{\left<k\right>} \sum_k 
\frac{\beta_1 \beta_3 \Theta^*_3 k^2 p_k}
{\beta_1+(\beta_1+1)\beta_3\Theta^*_3 k}.
\label{eq:schinazi99_jsp_h}
\end{equation}
As a result,
$\{0, 1, 3\}$ and $\{1\}$ are separated by
\begin{equation}
\beta_3 = \frac{\left<k\right>}{\left<k^2\right>}.
\end{equation}
In summary, 
the 
$\{0, 1, 2, 3\}$ phase dominates even for
infinitesimally small infection rates
when $\left<k^2\right> = \infty$ and $\beta_2 /(r+\phi)>\beta_3$.
When the latter condition is violated,
$\{0, 1, 3\}$ appears for a tiny infection rate.

\subsection{Superspreader Model}\label{sub:example3}

{\bf Model}

Superspreaders, namely, patients that infect many others
in comparison with normal patients, are identified in the outbreaks
in, for example,
gonorrhea \cite{Hethcote84}, HIV \cite{Anderson86,May88TRANS,Colgate},
and SARS,
\cite{Abdullah,Singapore}.
Such heterogeneous infection rates are sometimes ascribed to
heterogeneous social or sexual contact rates as specified by $p_k$
\cite{Hethcote84,Anderson86,May88TRANS,Liljeros,Meyers03,Schneeberger}.
However, superspreaders exist even in diseases whose associated
networks are considered to have relatively homogeneous $k$, such as
SARS \cite{Masuda_SARS}. Then, an alternative way to introduce
superspreaders is to consider two types of patients with different
infection rates. Then, competition occurs between superspreaders and
normal patients.

We assume that superspreaders and normally infectious patients are
not distinguished by the strain types.
Accordingly, an infection event originating from a
superspreader can cause a normal patient and vice
versa. There is no mutation, and competition occurs indirectly.
That is how the superspreader model considered here is
essentially different from the models in \SECS\ref{sub:example1} and
\ref{sub:example2}. 

\begin{figure}
\begin{center}
\includegraphics[height=2.25in,width=2.25in]{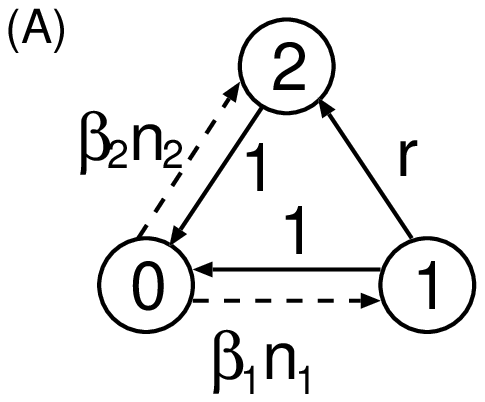}
\includegraphics[height=2.25in,width=3.25in]{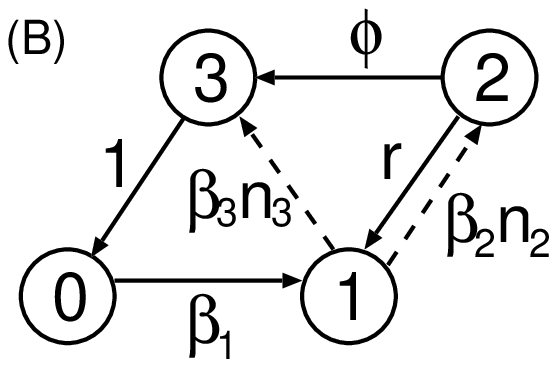}
\includegraphics[height=2.25in,width=4.00in]{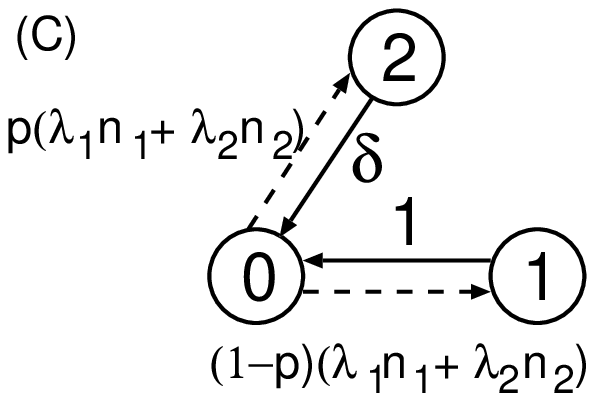}
\caption{The state-transition 
rules of the models with competing pathogens analyzed in 
(A) \SEC\ref{sub:example1}, (B) \SEC\ref{sub:example2}, and (C)
\SEC\ref{sub:example3}.}
\label{fig:twoout}
\end{center}
\end{figure}

\begin{figure}
\begin{center}
\includegraphics[height=2.25in,width=2.25in]{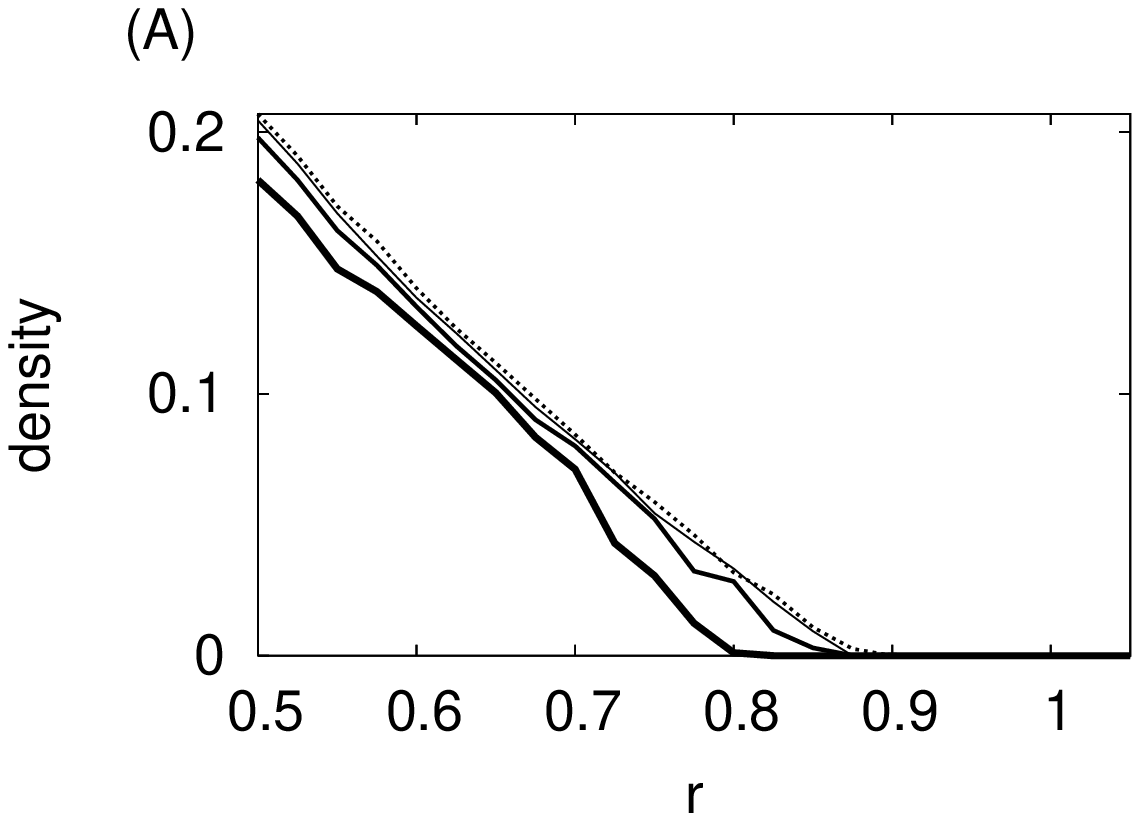}
\includegraphics[height=2.25in,width=2.25in]{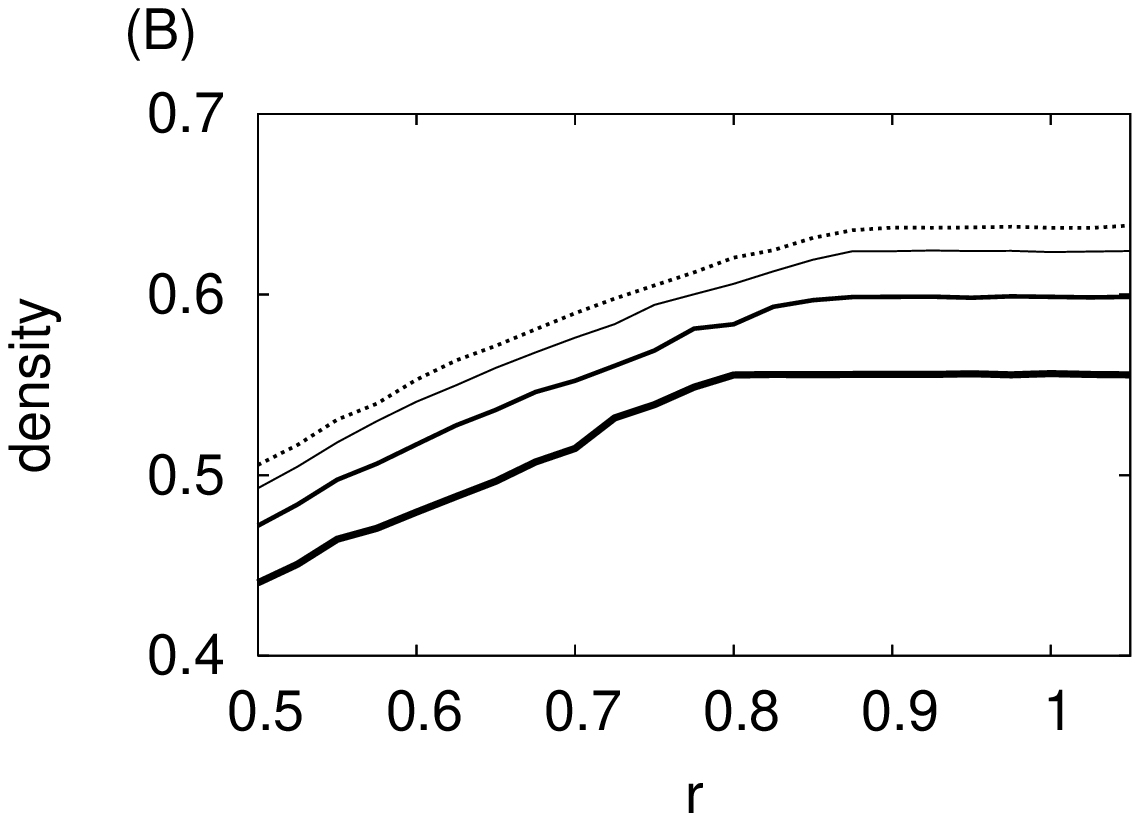}
\caption{Numerically obtained 
stationary densities of (A) state 1 and (B) state 2 for the model
in \SEC\ref{sub:example1}
(see \FIG\ref{fig:twoout}(A) for the transition rules).
The scale-free networks with $\gamma=2.5$
(thickest solid lines), $\gamma=3.0$ (moderate solid lines),
$\gamma=4.0$ (thinnest solid lines), and the random graph (dotted lines) 
are used. See the text for other parameter values.}
\label{fig:pathogen}
\end{center}
\end{figure}

We analyze a three-state model with different infection
rates, which is schematically shown in 
\FIG\ref{fig:twoout}(C) \cite{Kemper78,Schinazi01_MB}.
The states 0, 1, and 2
mean susceptible, patients of type 1 with infection rate $\lambda_1$
and recovery rate 1, and patients of type 2 with infection rate
$\lambda_2\ge \lambda_1$ and recovery rate $\delta$,
respectively. State 2 corresponds to the superspreader, and 
in practical appliations, we put $\lambda_1\ll \lambda_2$. 
Upon infection, a
state-0 vertex changes its state into 1 (2)
with probability $1-p$ ($p$).  There are two neighbor-dependent
infection routes in \FIG\ref{fig:twoout}(C). State 1 and state 2
indirectly compete with each other through crosstalk. In other words,
the density of state 1 and that of state 2
affect the rates of transitions $0\to 2$ and
$0\to 1$, respectively. According to the meanfield analysis,
$\{0, 1, 2\}$ results when
\begin{equation}
\lambda_1 \delta (1-p) + \lambda_2 p >\delta.
\end{equation}
Otherwise, $\{0\}$ results.
Slightly weaker but qualitatively similar results have been
proved for lattices \cite{Schinazi01_MB}.

{\bf Analysis}

In heterogeneous populations, dynamics are given by
\begin{eqnarray}
\dot{\rho}_{1,k} &=& (1-p) \left( 1-\rho_{1,k}-\rho_{2,k} \right)
k (\lambda_1 \Theta_1 + \lambda_2 \Theta_2)
- \rho_{1,k},\nonumber\\
\dot{\rho}_{2,k} &=& p \left( 1-\rho_{1,k}-\rho_{2,k}\right)
k (\lambda_1 \Theta_1 + \lambda_2 \Theta_2)
- \delta \rho_{2,k}.
\label{eq:schinazi01_mb_a}
\end{eqnarray}
The steady state is:
\begin{equation}
\left(\begin{array}{c}
\rho^*_{1,k}\\ \rho^*_{2,k}
\end{array} \right) =
\frac{(\lambda_1\Theta^*_1+\lambda_2\Theta^*_2)k}
{\delta+\left[p+\delta\left(1-p\right)\right]
(\lambda_1\Theta^*_1+\lambda_2\Theta^*_2)k}
\left(\begin{array}{c}
\delta(1-p)\\ p
\end{array} \right),
\label{eq:schinazi01_mb_b}
\end{equation}
which leads to
\begin{equation}
\Theta^*_1 = \frac{\delta(1-p)}{p}\Theta^*_2.
\label{eq:schinazi01_mb_c}
\end{equation}
Therefore, $\Theta^*_1>0$, or equivalently,
$\overline{\Theta}^*\equiv 
\lambda_1\Theta^*_1 + \lambda_2\Theta^*_2>0$, is
the condition for the $\{0, 1, 2\}$ phase.
With \EQ(\ref{eq:schinazi01_mb_b}),
we obtain 
\begin{equation}
\overline{\Theta}^* = \frac{1}{\left< k\right>}\sum_k
\frac{\left[\lambda_1\delta\left(1-p\right)+\lambda_2 p
  \right]\overline{\Theta}^* k^2 p_k}
{\delta+\left[p+\delta\left(1-p\right)\right]k\overline{\Theta}^*},
\label{eq:schinazi01_mb_d}
\end{equation}
and $\{0, 1, 2\}$ emerges when
\begin{equation}
\lambda_1 \delta (1-p) + \lambda_2 p >
\frac{\delta \left<k\right>}{\left<k^2\right>}.
\label{eq:schinazi01_mb_e}
\end{equation}
Equation~(\ref{eq:schinazi01_mb_e}) suggests that 
an arbitrary small infection rate $\lambda_1$ or $\lambda_2$ allows 
the endemic state as $\left<k^2\right>/\left<k\right>\to\infty$.
 In contrast to the models in
\SECS\ref{sub:example1} and \ref{sub:example2}, degree-independent
factor does not play a role in this model. This is
due to the absence of mutation from 1 to 2 or from 2 to 1.

\section{Double Infection Model}\label{sec:double}

{\bf Model}

For a realistic network model of
populations with natural birth,
we have to take into account that birth can occur only at empty
sites. Then, a minimal model such as the contact process
must be extended to a three-state process in which states 0, 1, and 2
mean empty, susceptible, and infected, respectively.  A susceptible
gives birth to an offspring in a neighboring empty vertex, 
which is operationally similar to a contagion event.
An infected preys on a susceptible in its neighborhood. Then there are two
neighbor-dependent transition rates, namely, 0 $\to$ 1 transition
at a rate proportional to $n_1$ and 1 $\to$ 2 transition at a rate
proportional to $n_2$.
A similar situation arises in the context of double infection in which
states 0, 1, and 2 correspond to susceptible, infected,
and infected by another pathogen, respectively.  A second pathogen
(state 2) targets individuals infected by a first pathogen
(state 1).

Keeping these interpretations in mind, we analyze the model shown in
\FIG\ref{fig:double} \cite{Sato94,Haraguchi}.  State 2 may more
virulent than state 1, and evolution of strains has been investigated with
a more complicated version of this model \cite{Haraguchi}.  The model
extends the contact process and the SIR model, which correspond to
$\mu = 0$ and $\lambda = \delta = r = 0$, respectively.

This model is theoretically intricate.
A subtlety stems from the cascaded neighbor-dependent
transitions 0 $\to$ 1 and 1 $\to$ 2 \cite{DurrettNeuhauser91,Andjel}.
There are three basic phases $\{0\}$, $\{0,1\}$,
and $\{0, 1, 2\}$.  The meanfield solution predicts that $\{0\}$ and
$\{0, 1\}$ are divided by $\lambda = \delta$. The one-dimensional
lattice with $r=0$ does not permit $\{0, 1, 2\}$
\cite{Sato94,Andjel}. However, on regular lattices of any dimension,
introduction of a small recovery rate $r>0$ elicits $\{0,1,2\}$
\cite{KST04}.  It is counterintuitive in the sense that $r$ is the
rate at which state 2 turns into state 1. Indeed, the meanfield theory
predicts the $\{0, 1, 2\}$ phase for sufficiently small $r$ fulfilling
\begin{equation}
\frac{\delta}{\lambda} + \frac{r+1}{\mu} <1.
\label{eq:kst-012-mean}
\end{equation}

Another paradoxical behavior occurs regarding to population density.
For example, an increase in $\lambda$ lessens the number of 
state-1 vertices on lattices because a
larger $\lambda$ creates more preys to devour for state-2 vertices
\cite{Tainaka03}.  Similarly, too large $\mu$ drives state 2 to perish
due to excess mortality \cite{Sato94,Haraguchi}, which
defines another $\{0, 1\}$ phase that we do not examine here.
This $\{0, 1\}$ phase is distinct from the $\{0, 1\}$ phase revealed by the
meanfield equations and mathematical analysis on lattices.

This model is complex also in the sense that 
the phase diagram in the parameter space is even qualitatively unknown
for lattices. The meanfield solution indicates that the two critical
lines approach each other as $\mu\to\infty$.  On the
other hand, they may cross at finite $\lambda$ and $\mu$, as supported
by the improved pair approximation ansatz \cite{Sato94,Haraguchi}.

{\bf Analysis}

In heterogeneous populations, the dynamics
read
\begin{eqnarray}
\dot{\rho}_{1,k} &=& \lambda (1-\rho_{1,k}-\rho_{2,k}) k
\Theta_1 - \delta \rho_{1,k} + r \rho_{2,k}
- \mu\rho_{1,k}k\Theta_2,\nonumber\\
\dot{\rho}_{2,k} &=& \mu\rho_{1,k}k\Theta_2  - (r+1)\rho_{2,k}.
\label{eq:schinazi04_mprf_a}
\end{eqnarray}
The steady state is given by
\begin{equation}
\left(\begin{array}{c}
\rho^*_{1,k}\\ \rho^*_{2,k}
\end{array}\right)
= \frac{\lambda\Theta^*_1 k}
{\delta(r+1)+\left[\lambda(r+1)\Theta^*_1+\mu\Theta^*_2\right]k
+\lambda\mu\Theta^*_1\Theta^*_2 k^2} 
\left(\begin{array}{c}
r+1\\
\mu \Theta^*_2 k
\end{array}\right),
\label{eq:schinazi04_mprf_b}
\end{equation}
which leads to
\begin{equation}
\left(\begin{array}{c}
\Theta^*_1\\ \Theta^*_2
\end{array}\right)
= \frac{1}{\left<k\right>}\sum_k
\frac{\lambda\Theta^*_1 k^2 p_k}
{\delta(r+1)+\left[\lambda(r+1)\Theta^*_1+\mu\Theta^*_2\right]k
+\lambda\mu\Theta^*_1\Theta^*_2 k^2} 
\left(\begin{array}{c}
r+1\\
\mu \Theta^*_2 k
\end{array}\right).
\label{eq:schinazi04_mprf_c}
\end{equation}
Clearly, $(\Theta^*_1, \Theta^*_2) = (0, 0)$,
which corresponds to the $\{0\}$ phase, solves
\EQ(\ref{eq:schinazi04_mprf_c}). To explore other phases, let us set
\begin{equation}
f_1(\Theta^*_1, \Theta^*_2) \equiv
\frac{\lambda(r+1)}{\left<k\right>}\sum_k
\frac{k^2 p_k}
{\delta(r+1)+\left[\lambda(r+1)\Theta^*_1+\mu\Theta^*_2\right]k
+\lambda\mu\Theta^*_1\Theta^*_2 k^2} - 1,
\label{eq:schinazi04_mprf_d}
\end{equation}
and
\begin{equation}
f_2(\Theta^*_1, \Theta^*_2)\equiv
\frac{\lambda\mu\Theta^*_1}{\left<k\right>}\sum_k
\frac{k^3 p_k}
{\delta(r+1)+\left[\lambda(r+1)\Theta^*_1+\mu\Theta^*_2\right]k
+\lambda\mu\Theta^*_1\Theta^*_2 k^2} - 1.
\label{eq:schinazi04_mprf_e}
\end{equation}
Then \EQ(\ref{eq:schinazi04_mprf_c}) is equivalent to
$\Theta^*_1 f_1(\Theta^*_1, \Theta^*_2) = 0$ and 
$\Theta^*_2 f_2(\Theta^*_1,
\Theta^*_2) = 0$.

First, we identify the boundary between $\{0\}$ and 
$\{0, 1\}$. In the $\{0, 1\}$ phase, we have
$\Theta^*_2 = 0$. Under this condition, we look for
$0<\Theta^*_1< 1$ that satisfies $f_1(\Theta^*_1, 0) = 0$.
By substituting $\Theta^*_2 = 0$ into
\EQ(\ref{eq:schinazi04_mprf_d}), we obtain 
\begin{equation}
f_1(\Theta^*_1, 0) = \frac{\lambda}{\left<k\right>}
\sum_k \frac{k^2 p_k}{\delta+\lambda \Theta^*_1 k} - 1 =
0.
\label{eq:schinazi04_phase01}
\end{equation}
Because
\begin{equation}
f_1(1, 0) =
\frac{\lambda}{\left<k\right>}\sum_k\frac{k^2 p_k}
{\delta+\lambda k} - 1
\le
\frac{\lambda}{\left<k\right>}\sum_k\frac{k^2 p_k}
{\lambda k} - 1\le 0,
\label{eq:schinazi04_mprf_d3}
\end{equation}
\EQ(\ref{eq:schinazi04_phase01}) is satisfied when
$f_1(0, 0)>0$, that is,
\begin{equation}
\frac{\lambda}{\delta} > \frac{\left<k\right>}{\left<k^2\right>}. 
\label{eq:lambdac_01}
\end{equation}
This conclusion complies with the results for the
contact process (\SEC\ref{sec:introduction}).

\begin{figure}
\begin{center}
\includegraphics[height=2.25in,width=2.25in]{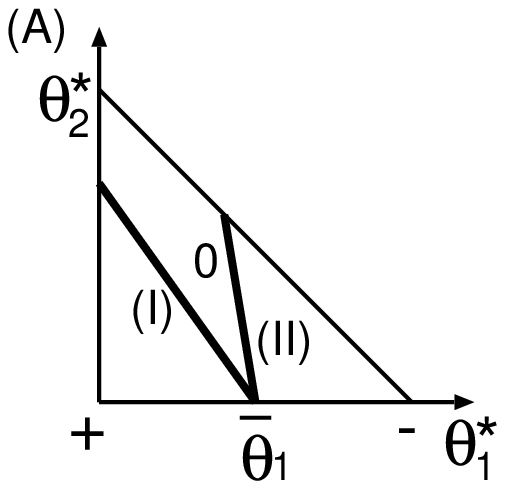}
\includegraphics[height=2.25in,width=2.25in]{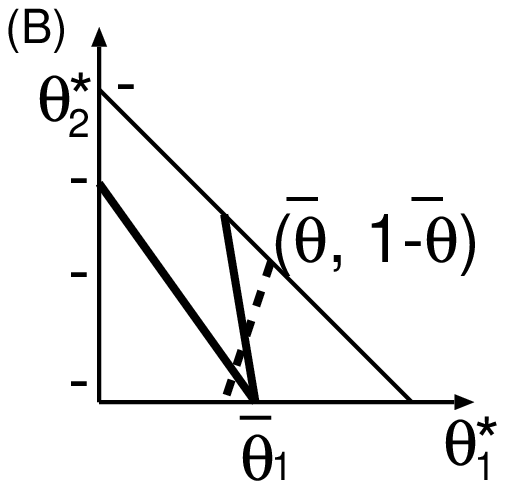}
\caption{(A) The isocline $f_1=0$ (solid lines) and sign of $f_1$, (B)
$f_2=0$ (dashed line) and sign of $f_2$ superimposed on
$f_1=0$ (solid lines).}
\label{fig:f1f2}
\end{center}
\end{figure}

Second, we examine the boundary between $\{0, 1\}$ and $\{0, 1, 2\}$.
The $\{0, 1, 2\}$ phase implies
$\Theta^*_1>0$ and $\Theta^*_2>0$ such that $f_1(\Theta^*_1,
\Theta^*_2) = 0$ and $f_2(\Theta^*_1, \Theta^*_2) = 0$. Let us suppose
that \EQ(\ref{eq:lambdac_01}) is satisfied because $\{0\}$ results
otherwise.  Then, there is a $0<\overline{\Theta}_1\le 1$ that
satisfies \EQ(\ref{eq:schinazi04_phase01}). Since $\partial f_1/
\partial \Theta^*_1 < 0$ and $\partial f_1/ \partial \Theta^*_2 < 0$,
there is a curve $f_1(\Theta^*_1, \Theta^*_2) = 0$ in the
$\Theta^*_1$-$\Theta^*_2$ space that looks like the solid line (I)
(when $f_1(0,1)\le0$), or (II) (when $f_1(0,1)>0$) in
\FIG\ref{fig:f1f2}(A).

Because $\partial f_2/ \Theta^*_1 > 0$ and $\partial f_2/
\partial\Theta^*_2 < 0$, $f_2$ monotonically increases in $\Theta^*_1$
on lines $\Theta^*_2 = 0$ and $\Theta^*_1+\Theta^*_2=1$.
If the isocline 
$f_2(\Theta^*_1, \Theta^*_2) = 0$ is located as the dashed line in 
\FIG\ref{fig:f1f2}(B), it nontrivially crosses
$f_1(\Theta^*_1, \Theta^*_2)
= 0$ (one of the solid lines) to yield
the $\{0, 1, 2\}$ phase.
Because $f_2(0, \Theta^*_2) =-1$ ($0\le \Theta^*_2
\le 1$), $f_2(1,0)>0$ is necessary for the dashed line to
exist. Then,
there is a unique $0<\overline{\Theta}<1$ so that
$f_2(\overline{\Theta}, 1 - \overline{\Theta}) = 0$.  Finally,
$f_1(\Theta^*_1,\Theta^*_2)=0$ and $f_2(\Theta^*_1,\Theta^*_2)=0$
cross when (i) $f_2(\overline{\Theta}_1, 0) > 0$ and (ii)
$f_1(\overline{\Theta},1-\overline{\Theta}) < 0$ are both satisfied.
Using 
\begin{equation}
f_1(\overline{\Theta}_1,0) = \frac{\lambda}{\left<k\right>}
\sum_k\frac{k^2 p_k}{\delta+\lambda\overline{\Theta}_1 k} - 1 = 0,
\label{eq:schinazi04_mprf_overline_theta1}
\end{equation}
the condition (i) is equivalent to
\begin{eqnarray}
f_2(\overline{\Theta}_1, 0) &=& 
\frac{\lambda\mu\overline{\Theta}_1}{(r+1)\left<k\right>}
\sum_k \frac{k^3 p_k}{\delta + \lambda\overline{\Theta}_1 k} - 1
\nonumber\\
&=& 
\frac{\lambda\mu\overline{\Theta}_1}{(r+1)\left<k\right>}
\sum_k \frac{k^3 p_k}{\delta + \lambda\overline{\Theta}_1 k} - 1
+ \frac{\mu\delta}{(r+1)\lambda}
f_1(\overline{\Theta}_1,0)\nonumber\\
&=& \frac{\mu\left<k^2\right>}{(r+1)\left<k\right>} - 
\left( 1 + \frac{\mu\delta}{(r+1)\lambda}\right)>0,
\label{eq:double-012a}
\end{eqnarray}
that is,
\begin{equation}
\frac{\delta}{\lambda} + \frac{r+1}{\mu} <
\frac{\left<k^2\right>}{\left<k\right>},
\end{equation}
which is an extension of the ordinary meanfield
solution (\EQ(\ref{eq:kst-012-mean})).
We note that (i) implies $f_2(1,0)>0$, which we required beforehand.

To show (ii), define
\begin{equation}
C_5\equiv \frac{\lambda(r+1)(1-\overline{\Theta})}
{\lambda(r+1)\overline{\Theta}+\mu(1-\overline{\Theta})}.
\end{equation}
We obtain
\begin{eqnarray}
&& f_1(\overline{\Theta},1-\overline{\Theta})\nonumber\\
&=& f_1(\overline{\Theta},1-\overline{\Theta})
+ C_5
f_2\left(\overline{\Theta},1-\overline{\Theta}\right)\nonumber\\
&=& 
\frac{\lambda}{\left<k\right>}\sum_k
\frac{(r+1+ C_5 \mu \overline{\Theta} k)
k^2 p_k}{\delta(r+1)+\left[\lambda(r+1)\overline{\Theta}+\mu
\left(1-\overline{\Theta}\right)\right]k
+\lambda\mu\overline{\Theta}\left(1-\overline{\Theta}\right)k^2}
- 1 - C_5\nonumber\\
&<& 
\frac{\lambda}{\left<k\right>}\sum_k
\frac{(r+1 + C_5 \mu \overline{\Theta}k)
k^2 p_k}{\left[\lambda(r+1)\overline{\Theta}+\mu
\left(1-\overline{\Theta}\right)\right]k
+\lambda\mu\overline{\Theta}\left(1-\overline{\Theta}\right)k^2}
- 1 - C_5\nonumber\\
&=& -\frac{\mu(1-\overline{\Theta})}{\lambda(r+1)
\overline{\Theta}+\mu(1-\overline{\Theta})} \le 0.
\label{eq:proof_r0}
\end{eqnarray}
To summarize, both $\{0\}$ and $\{0, 1\}$ disappear to
be replaced by $\{0,1,2\}$ as $\left<k^2\right>\to\infty$.

\section{Rock-Scissors-Paper Game}\label{sec:rock}

{\bf Model}

Winnerless cyclic competition among different phenotypes abounds in
nature.  For example, real microbial communities of {\it Escherichia
coli} \cite{Kerr02,Kirkup04} and color polymorphisms of natural
lizards \cite{Sinervo96} have cyclically dominating three states. The
evolutionary public-good game with volunteering (choice of not joining
the game) also defines a three-state population dynamics with cyclic
competition \cite{Szabo02,Semmann03}.  A typical consequence
of these dynamics is oscillatory
population density with each
phenotype alternatively dominant.

\begin{figure}
\begin{center}
\includegraphics[height=2.25in,width=2.25in]{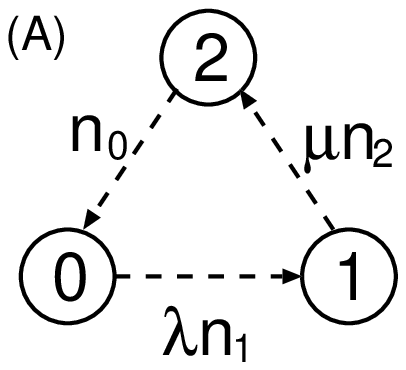}
\includegraphics[height=2.25in,width=2.25in]{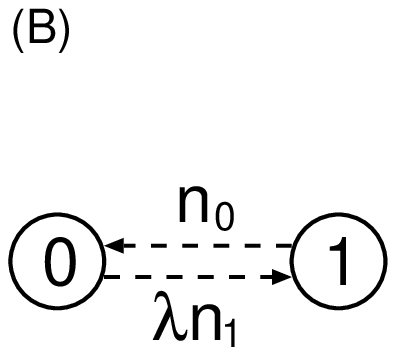}
\caption{The state-transition
rules of (A) the rock-scissors-paper game and
(B) the voter model.}
\label{fig:rock_voter}
\end{center}
\end{figure}

Let us consider a simple
rock-scissors-paper game with
cyclic competition (\FIG\ref{fig:rock_voter}(A)).
The ordinary meanfield analysis yields a single phase:
a neutrally stable periodic orbit in which the densities of 
states 0, 1, and 2 are cyclically dominant
\cite{Hopbauerbook}. The coexistence equilibrium inside the 
limit cycle is stabilized on 
regular lattices \cite{Tainaka94,Frean01}
and trees \cite{Sato97}. Consequently,
convergence to $\{0, 1, 2\}$
with a damped oscillation occurs
on these graphs.
We focus on the steady states only in the following analysis.

{\bf Analysis}

The dynamics in heterogeneous populations are given by
\begin{eqnarray}
\dot{\rho}_{1,k} &=& \lambda (1-\rho_{1,k}-\rho_{2,k}) k
\Theta_1 - \mu \rho_{1,k} k \Theta_2,\nonumber\\
\dot{\rho}_{2,k} &=& \mu \rho_{1,k} k
\Theta_2 - \rho_{2,k} k (1-\Theta_1-\Theta_2),
\label{eq:rock_a}
\end{eqnarray}
which yields
\begin{equation}
\left(\begin{array}{c}
\rho^*_{1,k}\\ \rho^*_{2,k}
\end{array}\right)
= \frac{\lambda \Theta^*_1}{(\lambda\Theta^*_1+\mu\Theta^*_2)
(1-\Theta^*_1-\Theta^*_2)+\lambda\mu\Theta^*_1\Theta^*_2}
\left(\begin{array}{c}
1-\Theta^*_1-\Theta^*_2\\
\mu\Theta^*_2
\end{array}\right).
\label{eq:rock_b}
\end{equation}
Noting that \EQ(\ref{eq:rock_b}) is independent of
$k$, we obtain
\begin{equation}
\left(\begin{array}{c}
\Theta^*_1\\ \Theta^*_2
\end{array}\right)
= \left(\begin{array}{c}
\rho^*_{1,k}\\ \rho^*_{2,k}
\end{array}\right)
= \frac{1}{\lambda+\mu+1}
\left(\begin{array}{c}
1\\ \lambda
\end{array} \right),
\end{equation}
for any $k$. The degree distribution does not affect the steady state.

Similarly, it is straightforward to show that
degree distributions are irrelevant in
the voter model (\FIG\ref{fig:rock_voter}(B)) and cyclic interaction
models with more than three states \cite{Sato02,Szabo04PRE}.
A lesson from these examples is that heterogeneous contact rates
do not influence the equilibrium population density
if there is no neighbor-{\it in}dependent state transition.
However, models with at least one neighbor-independent transitions 
show extinction of the epidemic threshold as
$\left<k^2\right>/\left<k\right>\to\infty$. 
For example, variants of the
rock-scissors-paper game supplied with spontaneous
death or mutation rates \cite{Tainaka03} are
expected to belong to the class analyzed in \SEC\ref{sec:double}.

\section{Two-population Model}\label{sec:malaria}

{\bf Model}

In many infectious diseases of humans, such as malaria, 
yellow feber, and dengue feber,
transmission are mediated by other hosts such as mosquitoes.
In these diseases,
direct human-human or mosquito-mosquito infection is absent, and
two distinct populations transmit diseases between each other.
As an example, we analyze a model for malaria spreads
(Anderson and May, 1991, Ch. 14).
Following the original notation, the
meanfield dynamics are represented by
\begin{eqnarray}
\dot{y} &=& \frac{ab\Omega}{N}(1-y)\psi - \gamma y,\nonumber\\
\dot{\psi} &=& acy(1-\psi)-\mu\psi,
\label{eq:malaria_mf_a}
\end{eqnarray}
where $y$ and $\psi$ are the proportions of infected
humans and mosquitoes, respectively. The size of the human population
is denoted by $N$, and $\Omega$ is the size of the female mosquito
population; only female mosquitoes infect humans. In addition, $a$ is
the bite rate, $b$ and $c$ are infection rates, $\gamma$ is the
recovery rate of the humans, and $\mu$ is the death rate of the
mosquitoes. Infected humans and infected
mosquitoes survive
simultaneously if they do. Based on
\EQ(\ref{eq:malaria_mf_a}), this occurs when 
\begin{equation}
R_0 \equiv \frac{a^2 bc\Omega}{N\gamma\mu}>1.
\label{eq:malaria_mf_b}
\end{equation}
Otherwise, the malaria is eradicated.

{\bf Analysis}

How does this condition change by introducing heterogeneous contact
rates? In network terminology, 
relevant networks are bipartite, with each part representing
the human population and the mosquito population.
Let us denote the degree distributions of humans
and mosquitoes by $\{p_{y, k}\}$ and $\{p_{\psi, k}\}$, respectively.
Real data do not suggest that
$\{p_{y, k}\}$ or $\{p_{\psi, k}\}$ is scale-free. 
However, we treat general degree distributions because
the obtained results may apply to
other multi-population contagion processes.

The dynamics are given by
\begin{eqnarray}
\dot{y}_k &=& \frac{ab\Omega}{N}(1-y_k) k \Theta_{\psi} - \gamma
y_k,\nonumber\\ 
\dot{\psi}_k &=& ac (1-\psi_k) k\Theta_y - \mu\psi_k,
\label{eq:malaria_a}
\end{eqnarray}
where $y_k$ ($\psi_k$) is the
probability that humans (mosquitoes) with degree $k$ are infected,
and
\begin{eqnarray}
\Theta_y &\equiv& \frac{1}{\left<k\right>_y}
\sum_k k p_{y, k} y_k,\quad \left<k\right>_y \equiv \sum_k k p_{y, k},
\nonumber\\
\Theta_{\psi} &\equiv& \frac{1}{\left<k\right>_{\psi}}
\sum_k k p_{\psi, k} \psi_k,\quad \left<k\right>_{\psi} \equiv
\sum_k k p_{\psi, k}.
\label{eq:malaria_b}
\end{eqnarray}
The steady state is calculated as
\begin{equation}
\left(y^*_k, \psi^*_k\right) = 
\left(
\frac{ab\Omega \Theta^*_{\psi}k}{N\gamma+ab\Omega \Theta^*_{\psi}k},
\frac{ac\Theta^*_y k}{\mu+ac\Theta^*_y k}
\right),
\label{eq:malaria_c}
\end{equation}
which is solved trivially by 
$(\Theta^*_y, \Theta^*_{\psi}) = (0, 0)$.
To explore nontrivial solutions, let us eliminate $\Theta^*_{\psi}$
from \EQS (\ref{eq:malaria_b})
and (\ref{eq:malaria_c}) to obtain
\begin{equation}
\Theta^*_y = \frac{1}{\left<k\right>_y}
\sum_k
\frac{ab\Omega k^2 p_{y, k}
\sum_{k^{\prime}} \frac{ac\Theta^*_y k^{\prime 2}p_{\psi, k^{\prime}}}
{\mu+ac\Theta^*_y k^{\prime}}}
{N\gamma\left<k\right>_{\psi}+ab\Omega k p_{y, k}
\sum_{k^{\prime}} \frac{ac\Theta^*_y k^{\prime 2}p_{\psi, k^{\prime}}}
{\mu+ac\Theta^*_y k^{\prime}}}.
\label{eq:malaria_d}
\end{equation}
The RHS of \EQ(\ref{eq:malaria_d})
is less than 1 when $\Theta^*_y = 1$.
The endemic state results if
\begin{equation}
\frac{\partial}{\partial \Theta^*_y}
\left(\mbox{RHS of \EQ(\ref{eq:malaria_d})}\right)\bigg|_{\Theta^*_y=0}
=
\frac{a^2 bc\Omega\left<k^2\right>_y \left<k^2\right>_{\psi}}
{N\gamma\mu\left<k\right>_y \left<k\right>_{\psi}} > 1,
\label{eq:malaria_e}
\end{equation}
which extends \EQ(\ref{eq:malaria_mf_b}).
Equation~(\ref{eq:malaria_e}) indicates that
divergence of just either
 $\left<k^2\right>_y$ or $\left<k^2\right>_{\psi}$
is sufficient for the epidemic threshold to disappear.
Even if both moments are finite, their effects are
multiplicative.
This is consistent with
the results for two-sex models of heterosexual HIV transmissions
\cite{May88TRANS} and 
the bond percolation
on bipartite graphs \cite{Newman02_bond,Meyers03}.

\section{Conclusions}\label{sec:conclusions}

We have analyzed various models of endemic infectious diseases in
populations with heterogeneous contact rates, or on complex
networks. The effects of heterogeneity on epidemic
thresholds are summarized in \TAB\ref{tab:summary}.  In many models,
diverging second moments of the degree distribution extinguishes the
epidemic thresholds, as reported previously for the percolation
\cite{Albert00NAT,Callaway00,Cohen00PRL,Newman02_bond}, the contact process
\cite{Hethcote84,Pastor01PRL,Pastor01PRE}, the SIR model
\cite{Anderson86,May88TRANS,May01,Moreno02EPB}, the SIRS model \cite{Liu04JSM},
and the household moodel \cite{Liu04PHA}. On scale-free
networks, which underly
 sexually transmitted diseases and computer viruses (see
 \SEC\ref{sec:introduction}
for references), introduction of a tiny amount of virus to a
population can cause an endemic
state even in models with more complex transition rules.

However, the models with competing pathogens and mutation
(\SECS\ref{sub:example1} and \ref{sub:example2}) show different
behavior. The heterogeneity in contact rates equally boosts the
infection strengths of competing pathogens in these models. Whether
multiple strains survive or one overwhelms the others
depends on network-independent
mutation rates that modulate relative strengths of
pathogens. The rock-scissors-paper game and the voter model
stipulate another class of models in which heterogeneity does not
affect the equilibrium population density at all. For the
heterogeneity to take effects, there must be at least one
neighbor-independent transition rate.

For a fixed graph or a contact rate distribution, one can create new
models that are not convered by this paper.  Our analysis has not been
exhaustive. However, we consider that what essentially matters is
gross arrangements of contagion pathways.  If there is at least one
transmission route independent of the neighbors' states, which is
not the case for the rock-srissors-paper game, the equilibrium
population density will
depend on degree distributions as does the contact process.
A care must be paid to cases of competing
pathogens with mutation. We hope that our results give prescription
for understanding complex-network consequences of other models.

A last note is on the stability of solutions.  For the contact process
on complex networks, the coexistence phase is stable if it exists
\cite{Andersonbook,Olinsky}. Similarly, the stability of the simplest
nontrivial phase is assured for models analyzed in this
paper. However,
stability analysis of other phases seems mathematically difficult. In
addition, the
stability of coexistence solutions of the rock-scissors-paper game
shows network dependence \cite{Masuda06RSP}.
These topics are warranted for future work.

\clearpage

\thispagestyle{empty}

\clearpage

\thispagestyle{empty}

\clearpage

\thispagestyle{empty}

\clearpage

\thispagestyle{empty}

\clearpage

\thispagestyle{empty}

\clearpage

\thispagestyle{empty}

\clearpage

\thispagestyle{empty}

\clearpage

\thispagestyle{empty}

\begin{table}
\begin{center}
\caption{Summary of the effects of heterogeneous contact rates
on epidemic thresholds.}
\begin{tabular}{|c|c|c|c|c|}
\hline
model & phases & threshold $\to 0$ & remarks & section \\
& & as $\left<k^2\right>\to\infty$  & & \\ \hline
CP& $\{0\}$, $\{0, 1\}$ & yes &  similar for SIR,
& \ref{sec:introduction}\\
& &  & SIRS, household & \\ \hline
two-stage CP & $\{0\}$,  $\{0, 1, 2\}$  & yes &  & 
\ref{sub:twostage}\\ \hline
tuberculosis & $\{0\}$, $\{0,1,2\}$ 
& yes &  & \ref{sub:tuber}\\ \hline
2 pathogens  & $\{0\}$, $\{0,2\}$,
& yes & $\{0,1,2\}$ requires a & 
\ref{sub:example1}\\
with mutation & $\{0,1,2\}$
& & $p_k$-indep. condition  & 
\\ \hline
drug-resistant & $\{1\}$, $\{0,1,3\}$,
& yes & $\{0,1,2,3\}$ requires a & 
\ref{sub:example2}\\
& $\{0,1,2,3\}$
& & $p_k$-indep. condition & 
\\ \hline
superspreader & $\{0\}$, $\{0,1,2\}$
& yes & & \ref{sub:example3}\\ \hline
double infection & $\{0\}$, $\{0,1\}$, $\{0,1,2\}$
& yes for both &  & \ref{sec:double}
\\
& 
& $\{0\}$ and $\{0,1\}$ &  &
\\ \hline
rock-scissors-paper & $\{0,1,2\}$  & no & similar for
& \ref{sec:rock}\\
&   &  & the voter model
& \\ \hline
malaria & nonendemic, endemic & yes &
& \ref{sec:malaria}\\ \hline
\end{tabular}
\label{tab:summary}
\end{center}
\end{table}

\end{document}